# Angle Dependent Interlayer Magnetoresistance (ILMR) in Multilayer Graphene Stacks

S. C. Bodepudi[1], Xiao Wang[2], S. Pramanik[1] (Corresponding author: spramani@ualberta.ca)

[1]Department of Electrical and Computer Engineering, University of Alberta, Edmonton, AB T6G 2V4, Canada

[2] School of Microelectronics and Solid State Electronics, UESTC, Chengdu, Sichuan, China

**Abstract.**

Interlayer magnetoresistance (ILMR) effect is explored in a vertical stack of weakly coupled multilayer graphene as grown by chemical vapor deposition (CVD). This effect has been characterized as a function of temperature and tilt angle of the magnetic field with respect to the interlayer current. To our knowledge, this is the first experimental report on angle dependent ILMR effect in graphitic systems. Our data agrees qualitatively with the existing theories of ILMR in multilayer massless Dirac Fermion systems. However, a sharper change in ILMR has been observed as the tilt angle of the magnetic field is varied. A physical explanation of this effect is proposed, which is consistent with our experimental scenario.

# I. Introduction.

Multilayer massless Dirac carrier systems have recently attracted significant attention due to a rich range of transport phenomena observed in these unconventional materials[1–7]. These materials are realized by stacking two-dimensional (2D) layers of massless Dirac carriers, while ensuring weak coupling between the neighbouring layers. Due to weak interlayer coupling, charge carrier motion is primarily confined in the 2D plane and the carriers in each layer obey linear energy dispersion. Thus these materials can also be viewed as "bulk 2D systems" with zero gap energy bands[2]. Current perpendicular to plane (CPP) or interlayer transport in these systems takes place via tunnelling between the weakly coupled layers and interlayer transport exhibits novel magnetoresistance (MR) effects that are not observed in other more conventional material systems.

The origin of this interlayer MR is intricately related to the existence of 2D massless Dirac carriers in individual layers. In presence of an out-of-plane magnetic field $B_z$, linear energy dispersion of each layer transforms into a series of Landau levels[8–10]. Of particular importance is the so-called "zero mode Landau level", which is a unique feature of 2D Dirac carriers[8–10]. This zero mode Landau level remains pinned at the Dirac point, which also coincides with the Fermi level where carrier occupancy probability is 1/2 [8–10]. Thus, interlayer carrier tunnelling now occurs between the zero mode Landau levels of the individual layers. As the magnetic field $B_z$ is increased, degeneracy of the zero mode Landau level and hence carrier density at zero mode increases, resulting in an increased tunnelling current and negative interlayer MR. As shown in ref.[1], interlayer resistance ($R_{zz}$) in this system has $1/|B_z|$ dependence and hence strong negative MR can be obtained for relatively small values of magnetic field. This effect is often dubbed "interlayer magnetoresistance" or, ILMR[1–3]. It is important to note that in this configuration, magnetic field ($B_z$) is parallel to the interlayer (CPP) current and hence no classical magnetoresistance effect is expected due to absence of Lorentz force. Further details of this ILMR effect is discussed in section II.

However, realization of such multilayer massless Dirac carrier systems is not straightforward. The material that has been extensively studied so far in this context is the organic compound α-(BEDT-TTF)$_2$ I$_3$ where BEDT-TTF represents bis(ethylenedithio)-tetrathiafulvalene[2,3]. This material has a multilayered structure in which conductive layers of (BEDT-TTF) molecules are separated by insulating layers of I$_3^-$. As a result, the conductive layers are weakly coupled to each other resulting in strong conductance anisotropy and strong 2D nature of the charge carriers. When subjected to high pressure (>

1.5 GPa)[2], carriers in the conducting layers follow Dirac-like energy dispersion. The ILMR effect described above has been reported in this system at low temperatures, ~ 10K[2,3].

Graphene, a single atomic layer of carbon atoms arranged in a honeycomb pattern[8,11,12], is a well-known 2D massless Dirac Fermion system. Existence of 2D Dirac carriers in graphene has been established experimentally via observation of unconventional half-integer quantum Hall effect[13,14]. In principle, multilayer massless Dirac carrier systems could be realized by stacking multiple graphene layers. In fact such a stacked configuration exists in nature and is commonly known as graphite. Unfortunately, the neighbouring graphene layers in graphite are generally strongly coupled, which results in 3D nature of the charge carriers instead of massless 2D behaviour. This is particularly true for most common phases of graphite such as Bernal (or AB) stacked and rhombohedral (or ABC) stacked graphite[15,16], which exhibit complex energy dispersion near the Fermi level instead of linear, massless dispersion of single layer graphene. As a result ILMR effect as described in ref.s[1–3] is rarely observed in graphitic systems[17,18].

However, interlayer coupling between neighbouring graphene layers in graphite can be significantly weakened if the neighbouring layers are misoriented with respect to each other so that AB or ABC stacking is destroyed. Such randomly oriented stack of graphene layers is often termed as "turbostratic graphite" and can be realized by chemical vapor deposition (CVD) technique[7,19–22]. Due to weak interlayer coupling in this system, charge carriers have 2D Dirac character, despite the presence of the neighbouring layers. Existence of 2D Dirac carriers in CVD grown multilayer graphene (MLG) has been demonstrated by various techniques such as Raman spectroscopy[15,19], scanning tunnelling microscopy[22], infrared spectroscopy[23] and even ab-initio calculations[24]. Thus, ILMR effect is expected to manifest in CVD grown multilayer graphene stacks (or, turbostratic graphite).

In our previous work[7] we considered CPP transport in MLG stack, CVD grown on Ni substrate. As grown graphene layers are misoriented, which has been confirmed by Raman spectroscopy and this observation is consistent with prior studies as well[19,22]. A very large negative CPP MR effect was observed, which persists even up to room temperature[7]. By means of control experiments it was shown that the observed negative CPP MR is due to the ILMR effect of the multilayered graphene stacks and not due to the Ni/graphene interface or the contacts[7]. In this study, however, only two orientations of the external magnetic field were considered: $B \parallel I$ ($\theta = 0°$) and $B \perp I$ ($\theta = 90°$), where $I$ is the interlayer (CPP) current. The observed MR effect was found to be consistent with the ILMR theory developed in ref.s[1,2]. However, ILMR has a unique dependence on the tilt angle $\theta$, which provides additional insight into interlayer magnetotransport of this unusual material system. The goal of this paper is to report angle dependent ILMR observed in multilayer graphene stacks, which hitherto has not been reported in literature. Unlike α-(BEDT-TTF)$_2$ I$_3$, ILMR in MLG survives up to much higher temperatures, is observed at smaller fields, and does not require application of external pressure. This effect is therefore promising for future development of graphene-based flexible magnetic sensors and data storage elements.

This paper is organized as follows. In section II we briefly review the key characteristics of the ILMR effect and its angular dependence. Next, we describe MLG device preparation and characterization in section III, which is followed by results and discussion in section IV. We summarize and conclude our work in section V.

**II. ILMR Effect.**

The basic idea of ILMR has been briefly outlined above and is schematically described in Figure 1. In this section, we review the prior works on ILMR[1–4] and highlight the key features of this effect. Most of this work has been performed on α-(BEDT-TTF)$_2$ I$_3$. As described before, a vertical stack of weakly coupled 2D Dirac carrier systems is considered, in which Fermi level resides in the vicinity of the Dirac points where the density of states is small (in absence of any external magnetic field, Figure 1a). Any interlayer charge transfer occurs via tunnelling between the states located close to the Fermi level. In this case, interlayer current is small, due to lack of available states (and low number of carriers) in the vicinity of the Fermi level. In this case, therefore large interlayer resistance ($R_{zz}$) is expected.

This situation changes dramatically when an out-of-plane magnetic field ($B_z$) is applied. The linear energy dispersion of 2D Dirac carriers now converts into a series of Landau levels (Figure 1b), given by $E_n^{LL} = \pm\sqrt{(ehv_F^2|n||B_z|/\pi)}$[1,8–10], where $v_F$ is the Fermi velocity and $n$ is an integer representing the Landau index. Most importantly, a "zero mode" Landau level corresponding to $n = 0$ exists at the (quasi) Fermi level (Figure 1b). This zero mode Landau level is a signature of 2D Dirac materials and as can be seen from the above expression of $E_n^{LL}$, its location is independent of the applied magnetic field. As the out-of-plane magnetic field is increased, degeneracy ($eB_z/h$) of the zero mode Landau level and carrier concentration of zero mode increases. Since interlayer current is carried by the charge carriers located in the vicinity of the (quasi) Fermi level (Figure 1b), increased magnetic field increases the interlayer current, which results in negative interlayer magnetoresistance (ILMR). As noted before, in this measurement configuration ($\theta = 0$, Figure 2a) magnetic field does not exert any Lorentz force on the charge carriers.

When the magnetic field is slightly tilted from the out-of-plane direction ($\theta \neq 0$, Figure 2a), Lorentz force on the charge carriers is non-zero and carrier trajectory deflects from the out-of-plane direction. As a result, effective tunnelling distance between neighbouring layers increases, resulting in smaller interlayer current or reduced ILMR. In the limit of in-plane magnetic field, interlayer current is weakest and a large interlayer resistance is observed.

This physical picture has been modeled in ref.[1], which derived the following expression for interlayer resistivity ($\rho_{zz}$) under dc bias and in presence of an external magnetic field $\boldsymbol{B}$ ($B_x$, $B_y$, $B_z$):

$$\boldsymbol{\rho_{zz}(B)} = \frac{\pi \hbar^3}{2C\tau t_c^2 ce^3} \frac{1}{|B_z|} exp\left[\frac{1}{2} \frac{ec^2(B_x^2 + B_y^2)}{\hbar |B_z|}\right] \quad (1)$$

where $\tau$ is the characteristic life time (or relaxation time for in-plane scattering), $c$ is the interlayer spacing (~ 0.342 nm for turbostratic graphite[15]), $e$ is electronic charge, $\hbar$ (= $h/2\pi$) is reduced Plank constant and $t_c$ is the "interlayer transfer energy", which represents the degree of coupling between neighbouring graphene layers[1]. The condition of "weak interlayer coupling" requires $t_c$ to be smaller than disorder induced broadening ($h/2\pi\tau$) and thermal broadening ($k_B T$)[1]. For graphene, $h/2\pi\tau$ ~ 30K[25] (~ 3 meV) and for weakly coupled graphene layers, $t_c$ ~ 2meV[18,26]. Thus, the condition of "weak interlayer coupling" holds over our measured temperature range of 10K–200K.

According to ref.[1], $C$ ~ $1/k_B T$ for "high temperatures" i.e. $k_B T >> t_c$, $h/2\pi\tau$. As indicated above, in case of graphene, disorder induced broadening $h/2\pi\tau$ ~ 30K (or, ~ 3meV[25]) and high temperature limit can be attained for temperatures above 30K[7]. From the above expression, it is clear that for purely out of plane magnetic field ($B_x$, $B_y$ = 0), $\rho_{zz}$ ~ $1/|B_z|$, which is the origin of large negative ILMR. In the limit of purely in-plane field ($B_z$ = 0), $\rho_{zz}$ approaches infinity because the carriers will be strongly deflected towards the plane, which results in very low interlayer tunnelling probability due to large increase in effective tunnelling distance. Using the above expression of $\rho_{zz}$, the following angle ($\theta$) dependence can be obtained:

$$\rho_{zz}(\theta) = \frac{\pi\hbar^3}{2C\tau t_c^2 ce^3} \frac{1}{B|cos\theta|} exp\left[\frac{1}{2}\frac{ec^2 B sin^2\theta}{\hbar|cos\theta|}\right] \quad (2)$$

As can be seen from the above formula, for a given field strength $B$, $\rho_{zz}$ increases as the deviation ($\theta$) from the plane normal is increased. For $\theta = 90°$ and $270°$, equation 2 diverges. Physical meaning of this has been discussed above. In principle, this Lorentz force deflects the carriers in the in-plane direction and prevents them from reaching the other contact, resulting in an infinite interlayer resistance.

The above formulae (equations 1 and 2) for $\rho_{zz}$ assume that interlayer charge transport occurs between the zero mode Landau levels of the individual layers and no other level participates in conduction, which is the so-called "quantum limit"[1]. While this assumption is valid for the intermediate values of magnetic field, other effects come into play at smaller and higher field values. For example, if the magnetic field ($B_z$) is weak, separation between $n = 0$ and $n = 1$ levels ($\Delta E = \sqrt{(ehv_F^2|B_z|/\pi)} = \sqrt{(ehv_F^2|B\cos\theta|/\pi)}$) will be smaller than Landau level broadening ($\Gamma \sim \max(h/2\pi\tau, k_BT)$) and in this case both modes will participate in interlayer transport. It is straightforward to find out the "critical value" of the applied field $B$ (say, $B_{cr}$) above which "quantum limit" is achieved. For graphene, using $h/2\pi\tau \sim 30K$[25], we obtain $B_{cr} \sim 68G/\cos\theta$ for $T \leq 30K$. For higher temperatures, $\Gamma \sim k_BT$ and in this regime $B_{cr}(T) \sim T^2/\cos\theta$. Clearly, for $B < B_{cr}(T)$, Landau level mixing takes place and it has been shown in ref.[4] that such mixing leads to a positive magnetoresistance effect at low fields *due to non-vertical tunnelling processes*. Thus, for $B \sim B_{cr}(T)$, a crossover from positive to negative MR takes place and the expression for $\rho_{zz}(B)$ mentioned above (equation 1) ceases to be valid for $B < B_{cr}(T)$.

At high magnetic field limit, zero mode Landau level will be Zeeman split, resulting in reduced number of available states in the vicinity of the (quasi) Fermi level, which gives rise to a crossover from negative to positive magnetoresistance[1]. For this effect to take place in MLG, Zeeman splitting energy ($g\mu_B B_z \sim 0.12 B_z$ meV[8], $\mu_B$ being the Bohr magneton and $B_z$ measured in Tesla) must exceed broadening (thermal and disorder induced) of the zero mode Landau level. Even in the low temperature limit, where Landau level broadening is $\sim 30K$ as mentioned before, observation of the above effect will require a magnetic field of $\sim 25T$, which is beyond our measurement range. So the positive magnetoresistance effect due to Zeeman splitting is unlikely to occur in the present study.

It is important to note that the above-mentioned physical picture of ILMR remains valid even when the Fermi level resides within some higher order Landau level ($n > 0$). Effect of higher order Landau levels have been studied in ref.[4] and it has been found that $n \rightarrow n$ interlayer tunnelling leads to negative ILMR even when $n \neq 0$. As discussed before, at small fields, Landau levels overlap and $n \rightarrow n'$ ($n \neq n'$; $n, n' \neq$

0) interlayer tunnelling takes place. This process leads to positive MR at small fields for *non-vertical interlayer tunnelling*[4]. Since inter Landau level spacing decreases with increasing *n*, the low-field positive MR effect will become more dominant as *n* increases. Similarly, if the tilt angle $\theta$ is increased for a given field strength *B*, positive MR is expected to become even stronger. This is due to two reasons: first, tilted *B* results in reduced $B_z$ and reduced Landau level spacing, which leads to significant Landau level overlap and inter Landau level mixing. Second, tilted *B* will tend to deflect the carriers towards "in-plane" direction, thus increasing in-plane scattering and resulting in more "non-vertical" interlayer tunneling incidents. In addition, in-plane scattering processes themselves result in positive MR[27] (also see supplementary information[37]) and for $\theta \neq 0$ this effect is significant since carriers will experience significant in plane motion during interlayer transport. Such effects are not addressed by the model described by equations (1) and (2) and hence strong deviation is expected for $n \neq 0$ and $\theta \neq 0$.

Finally, it is important to note that the ILMR theory discussed above only models interlayer transport between the weakly coupled layers. In order to apply this theory to experimental results, it is important to establish that the measured resistance arises due to this effect and not from the contacts or interfaces or any other sources. As described below, we have confirmed this in our data analysis.

**III. Device Fabrication and Characterization.**

Figure 2(a) shows the schematic of the Ni/MLG/Ag device structure and the CPP measurement geometry. Device fabrication steps have been described in detail previously[7,28]. Briefly, polycrystalline Ni foils with primarily (111) crystal orientation are used as the catalyst for CVD growth of MLG. The Ni foil also serves as the bottom electrical contact for CPP measurement. In CVD process, polycrystalline nickel samples (2cm x 2cm) are first preheated in quartz tube up to 1000°C and then annealed in presence of hydrogen flow for one hour. Next, a precursor gas mixture, containing 0.2% $CH_4$, 9.8% $H_2$ and 90% Ar, is flown for 10 minutes and MLG growth takes place during this period. Finally, samples are naturally cooled (~ 3°C/min) to room temperature. As characterized in our previous studies[7,28], as-grown MLG on Ni consists of two distinct regions. The region (~ few layers) close to the graphene/Ni interface is generally defective due to strong overlap between 3*d* states of Ni and 2$p_z$ states of carbon[28]. In addition, the interfacial layers contain large number of atomic steps and grain boundaries[29–31]. The presence of defects has been confirmed by taking Raman spectrum from this region, which shows strong *D* (defective) peak (see later in Figure 3(b)). Graphene layers above this region are "defect-free" and these layers are weakly coupled to each other ("turbostratic")[7]. Again, these features have been confirmed by Raman studies as shown later in Figure 3(a). A schematic of the MLG stack on Ni substrate is shown in Figure 3(b) inset, which highlights the origin of grain boundaries in the interfacial layers and lack of these defects away from the bottom interface. Transferring the MLG layer on another substrate generally destroys the weak interlayer coupling (see later in Figure 3(c)) and hence CPP measurements have been performed on the as-grown samples[7].

To perform CPP-MR measurements on as-grown MLG on Ni (Figure 2(b), main image), we used silver (Ag) epoxy as the top contact, with contact area ~ 1mm$^2$ (Figure 2(a)). This contact is placed at the centre of the top MLG surface to ensure uniform current distribution. We also transferred the MLG on SiO$_2$/Si substrate using a previously reported[7,28] procedure to perform thickness and in-plane MR measurements. Optical images of the transferred MLG and its thickness distribution are presented in Figure 2(b), insets. Thickness measurements have been performed on the wrinkle-free areas of the transferred MLG and the average thickness is ~ 60nm.

To investigate the structural quality of as-grown MLG on Ni, we acquired Raman spectra using laser excitation of 532nm (2.33eV). Figure 3(a) shows the Raman spectra from three representative areas of as-grown MLG on Ni. Strong $G$ peak (1580 cm$^{-1}$) and absence of $D$ peak (1350 cm$^{-1}$) have been observed in all cases. This indicates formation of hexagonal lattice of carbon atoms without any significant structural defects[15]. It is to be noted that the penetration depth of Raman laser in graphite is ~50 nm[32] and hence this observation is valid for the graphene layers close to the top surface (the so-called "defect-free layer" in Figure 2(a)). The top plot (blue) in Figure 3(a) is most commonly observed (> 80% area). However, in all cases the position of the $2D$ band is ~ 2705 cm$^{-1}$ with line-width of ~ 60–80cm$^{-1}$ and the $2D$ band can be fitted with single Lorentzian peak. This is a typical signature of MLG with weak interlayer coupling ("turbostraticity")[7,15], which is a necessary prerequisite for observation of ILMR.

Figure 3(b) shows the Raman spectrum taken from the bottom MLG surface (i.e. MLG layers close to the Ni interface). To acquire this spectrum, Ni is first etched away and MLG is transferred on SiO$_2$/Si so that the bottom MLG surface faces up. A clear Raman $D$ peak has been observed, which confirms the defective nature of this region. The defect peak originates from the edges of small-area graphene sheets that form near Ni/MLG interface and defects created by hybridization between Ni 3$d$ and C 2$p_z$ orbitals. The graphene layers near this interface have smaller area because their growth starts "horizontally" from the Ni grain boundaries (instead of "vertically" from the Ni surface) and the planar geometrical shape of these graphene layers are determined by the grain boundary distribution of the Ni substrate. Thus, these layers contain numerous truncated graphene planes (Figure 3(b) inset), which contribute to the *D* peak in the Raman spectrum. However, away from this interface, grain-boundary growth sites are no longer available and graphene layers tend to grow "vertically" on top of the underlying layers. This forms a continuous, undulating coverage over the underlying discontinuous films (Figure 3(b) inset). As a result, the layers away from Ni/MLG interface are free from the edge states. Also, interfacial hybridization effects are absent away from the interface. Due to these reasons, *D* peak is absent in the Raman spectrum taken from the MLG stack away from the interface (Figure 3(a)). The dark contrast in Figure 2(b), *main image*, is due to the unevenness in layer thickness near the Ni/MLG interface. After transferring on a flat $SiO_2$ substrate, this uneven bottom surface creates a wrinkled appearance as shown in Figure 2(b), *left inset*. Details of graphene growth on Ni and complete evidence of the physical picture presented above are available in ref.s[29,30].

Figure 3(c) shows the Raman spectra taken from three representative areas of MLG after transferring on $SiO_2$/Si. Unlike the as-grown samples in Figure 3(a), the Raman 2*D* band of the transferred MLG consists of either a shoulder or a strong splitting, which is reminiscent of HOPG (highly oriented pyrolytic graphite) in which the graphene layers are primarily Bernal stacked[15]. The presence of shoulder or significant splitting in Raman 2*D* band is a signature of strong interlayer coupling in MLG[15] and this change in the 2*D* band behaviour originates due to the transfer process[7]. Thus for ILMR studies we have chosen as-grown MLG (on Ni) in which the graphene layers are weakly coupled. As reported in our previous work[7], transferred MLGs as characterized above do not show any ILMR effect.

We also note absence of *D* peak in these transferred samples (Figure 3(c)). This proves that the transfer process does not introduce any significant defect in the sample. Thus, the *D* peak observed in Figure 3(b) does not originate from the transfer process and indeed comes from the other sources as described above.

We have further characterized the transferred MLG using various electrical measurements (supplementary information (SI)[37]). Sheet resistance of transferred MLG (~ 60 nm thick) is ~ 50-100Ω, depending on temperature (SI[37], section 1). This is in good agreement with literature, where similar sheet resistance values were reported for Ni-grown MLG of similar thickness[33,34]. Typical contact resistance between MLG and Ag paste has been found to be ~ 3-10 Ω (SI[37], section 1), which is an order of magnitude smaller than the zero field CPP resistance reported in the next section. Thus MLG/Ag contact resistance does not play a crucial role in our experiments. Further, in-plane MR measurements performed on transferred MLG (SI[37], section 2) do not show any weak localization effect even at low temperatures. This is consistent with non-observation of defect peak in the Raman spectrum. In contrast, copper grown MLG shows presence of grain boundaries and defects, which are detected in the Raman spectrum (*D* peak) and as well as in the weak-localization feature in the planar MR measurements. Finally, in-plane MR measurements also reveal signatures of Shubnikov-deHaas (SdH) oscillations within 1T (SI[37], section 3), which indicates formation of Landau levels in this field range. From the periodicity of SdH oscillations, carrier concentration per layer is estimated to be ~ $10^{10}$/cm$^2$, which implies $n$ ~ 1-2 Landau levels are occupied at $B$ ~ 2kG. As described in the next section, this is roughly the field range where the negative MR manifests.

**IV. Results and Discussion.**

In this work, CPP resistance has been measured with a bias current of 1mA. Temperature dependence of CPP resistance shows insulating behaviour (Figure 4a), which is common for out of plane transport in disorder free graphite[35,36]. The zero field CPP resistance $R_{zz}(0)$ is ~ 190Ω at 15K and ~ 80Ω at 220K (Figure 4(a)). Application of a large magnetic field (8kG) results in a decrease in device resistance, but the insulating temperature dependence still persists (Figure 4(a)), indicating absence of any magnetic field induced metal-insulator transition effect. Also note that the $R_{zz}(0)$ values mentioned above are at least an order of magnitude higher than the MLG/Ag contact resistance (see SI[37], section 1). Thus MLG/Ag contact resistance does not play a significant role in the observed MR characteristics.

Figures 4(b), (c) show normalized CPP MR [$R_{zz}(B)/R_{zz}(B=0)$] of as-grown MLG on Ni in the field range ± 8kG for $\theta$ = 0° (or 180°) and 90° (or 270°). As evident from Figures 4(b), (c), the CPP resistance of as-grown MLG on Ni is strongly dependent on the direction of the magnetic field ($\theta$). When the magnetic field is normal to the graphene plane, i.e. $B \parallel I$ ($\theta$ = 0° or 180°), we observe ~ 40% drop in CPP resistance in the vicinity of 2kG (at 15K, Figure 4(b)). However, when the magnetic field is in-plane, i.e. $B \perp I$ ($\theta$ = 90°, 270°), we observed only a weak positive MR of ~ 8% (at 15K, within the measurement range of ± 8kG, Figure 4(c)). Such MR features cannot be explained with any semi-classical theory since Lorentz force on charge carriers is negligible when $B \parallel I$ and strongest for $B \perp I$. However, as described below, these MR characteristics are consistent with the ILMR picture discussed in section II.

One key feature of ILMR is the inverse dependence of CPP resistance ($R_{zz}$) on the out of plane component of the magnetic field ($B_z$). As discussed in section II, such dependence is only valid in the "intermediate field" range, where only a single mode (but not necessarily the zero mode) participates in interlayer transport and high field effects as well as low field Landau level mixing effects[4] are absent. In Figure 4(d) we have plotted inverse of (normalized) interlayer resistance as a function of normal component of the magnetic field in the range where negative MR is most prominent ($B > B_{cr}$). A clear linear fit is observed at all measurement temperatures as expected from theory. Figure 4(d) inset shows temperature dependence of the critical magnetic field ($B_{cr}$). As expected, based on the discussion in section II, $B_{cr} \propto T^2$, which is a signature of the ILMR effect. In this plot $B_{cr}$ is the value of the magnetic field at which device resistance starts to decrease and this quantity is closely related to the width (full width half maximum) of the MR curves in Figure 4(b). Clear linear fit in the $B_{cr}$-$T^2$ plot, and absence of any saturation at low $T$ implies that $\Gamma \propto k_B T$ even at the low $T$ limit. Thus disorder induced broadening is not significant in the present case, which is also consistent with our previous Raman characterization (Figure 3(a)).

As discussed above, MLG/Ag contact resistance does not play a major role in the CPP measurements. In our previous work[7] we showed that the metal contacts (Ni and Ag) or the interfacial regions are also not responsible for the observed negative CPP MR in these samples. For example, unlike the actual devices the metallic contacts themselves show metallic temperature dependence of resistivity. Further the metallic contacts have significantly smaller resistance than the actual device. Characterization of the "defective layer" at Ni/MLG interface also revealed metallic temperature dependence[7]. Thus the observed CPP MR originates from the graphene stacks away from the Ni/MLG interface, as a result of the ILMR effect. As characterized in Figure 3 by Raman spectroscopy, these layers are indeed defect free and weakly coupled, which are necessary prerequisites for observation of ILMR. As described later, zero field resistance $R_{zz}$ ($B = 0$) scales with MLG thickness, which further confirms that the CPP resistance originates from the "bulk" region of the MLG and not from the interfaces.

Figures 5 (a), (b) show the angular ($\theta$-dependent) response of CPP MR in MLG/Ni samples at 15K and 220K respectively. As consistent with the ILMR model described in section II, negative MR is strongest for $\theta = 0°$, 180° (**B** normal-to-plane) and gradually weakens as $\theta$ is tilted away from this direction. The measured CPP resistance curves for $\theta$, $180° \pm \theta$ and $360° - \theta$ are almost identical to each other, which is consistent with the expression of $\rho_{zz}$ ($\theta$) described in section II. Decrease in negative MR with increasing tilt angle ($\theta$) can be viewed as a result of a competition between the negative MR effect (due to $B_z$ component) and a positive MR effect (due to in-plane components $B_x$, $B_y$), which becomes stronger at larger $\theta$ and larger |B|. Additionally, as described before, a positive MR effect arises at low field range ($B < B_{cr}$) as well where inter Landau level mixing takes place[4]. However, negative MR effect generally dominates at higher fields ($B > B_{cr}$) for small $\theta$. Due to these competing effects, the MR characteristics tend to broaden as $\theta$ is increased, and often "shoulder"-like features are observed in the MR characteristics. As a result, $B_{cr}$ as defined earlier, increases with $\theta$. Physically, at a higher tilt angle, larger magnetic field needs to be applied to reduce Landau level overlap and overcome the resulting low field positive MR. In addition, increased in-plane scattering at higher tilt angle can further enhance the Landau level mixing effect described in ref.[4].

Figure 5(c) shows the angle dependence of the critical field ($B_{cr}$) at two different temperatures. As expected, at both temperatures critical field shows an increasing trend with the angular deviation ($\theta$). The negative MR and the interplay between positive and negative MR can be observed up to $\theta = 10°$ (Figures 5(a), (b)) and then the positive MR completely overwhelms negative MR. In Figure 5(c), $B_{cr}$ has been found to be higher for 220K as compared to 15K. This is expected due to the reasons described above. The error bars in Figure 5(c) originate due to the uncertainty in determining the tilt angle in our measurement setup. The tilt angle is measured using a needle attached to the rotatable sample holder and the width of the needle tip is ~ 1°, which introduces an uncertainty of +/- 1° in the angle measurements.

Figures 6(a), (b) show angle dependence of the normalized MR over the entire range of 0°–360° for three different values of (fixed) field strengths ($B > B_{cr}$) at two different temperatures. The data shows 180° periodicity and identical sharp dips for tilt angles $\theta = 0°$, 180° and 360°, both of which are consistent with the ILMR model described in section II. For a given field strength, device resistance increases as the tilt angle is increased with respect to the above-mentioned values. As described before, this increase arises from the Lorentz force due to the increased in-plane field component. However, the dips in the vicinity of these angles is sharper than that predicted by theory[1] (equations (1) and (2)), and device resistance tends to saturate at ~ 10° deviation from the above-mentioned angles. According to the theory[1], change in resistance with angle (at a given field strength) is more gradual (Figure 6(c)). The reason for this discrepancy can be understood as follows.

The theoretical model described in equations (1) and (2) predicts a positive MR as $\theta$ is increased. This model considers carrier deflection in presence of a tilted magnetic field and resulting increase in effective interlayer tunnelling distance (or reduced overlap between the wave functions of the neighbouring layers), which causes the positive MR for $B > B_{cr}$. However, there is another factor that contributes to the positive MR at higher fields. For example, tilted $B$ implies more in-plane carrier scattering in the current path, which results in a positive MR[27] (also SI[37], section 2). In a MLG stack containing ~ 100 layers or more, carriers undergo significant in-plane scattering in each layer in presence of tilted $B$ during interlayer transport, which can result in a significant net positive MR. However, this effect was not considered in equations (1) and (2), and as a result these equations underestimate the effect of the tilt angle on the observed MR.

As noted in Figure 4(b), MR at $\theta = 0^o$ (180$^o$) is smaller than that reported in our previous study[7]. This difference can be attributed to the thickness of the MLG considered or the number of graphene layers participating in interlayer transport. For thicker samples (~ 200 nm), such as in ref.[7], $R_{zz}$ ($B = 0$) is large (~ 650 Ω) due to large number of weakly coupled graphene layers along the out of plane direction. In this case, $R_{zz}$ ($B = 8kG$) is significantly small due to an abundance of carriers generated in a large number of graphene layers. As a result, very strong negative MR effect was observed in thicker MLGs, with a factor of ~ 160 drop in device resistance[7]. However, in the present case we have used thinner MLG (~ 60nm), which resulted in smaller zero field resistance $R_{zz}$ ($B = 0$) ~ 200 Ω. In presence of 8kG field, carriers will be generated but in fewer numbers due to fewer number of graphene layers. As a result, weaker MR effect will be observed for thinner samples, which is consistent with our observation. It is important to note that the zero field CPP resistance scales with MLG thickness for the same contact area, which confirms that zero field resistance originates from the "bulk" (or the "defect-free region") and not from the contacts or the interfaces.

## V. Summary and Conclusion.

In conclusion, we have reported experimental measurement of interlayer magnetoresistance effect in a vertical stack of randomly oriented graphene layers (or, turbostratic graphite). Presence of this effect confirms existence of two-dimensional massless Dirac carriers in this system. Temperature, field and angular dependences of this effect agree well with theory. The angular response is sharper than expected and is related to the additional sources of positive MR present in the system. In graphitic systems such as above, this effect persists at temperatures much higher than that reported for α-(BEDT-TTF)$_2$ I$_3$, presumably due to higher Fermi velocity in graphene (~10$^6$m/s, as opposed to ~10$^5$m/s for α-(BEDT-TTF)$_2$ I$_3$), which leads to larger separation between the Landau levels. At the same time, this also explains why this effect is observed at a much lower field (~ 0.2 T) in MLG as compared to α-(BEDT-TTF)$_2$ I$_3$ (~ 2T). Due to the strong MR signal at higher temperatures and lower field range, this effect is promising for next generation of flexible memory and sensor devices.

*Acknowledgement: This work has been funded by NSERC Discovery Grant. X.W. was supported by MITACS Globalink Summer Research program (2015).*

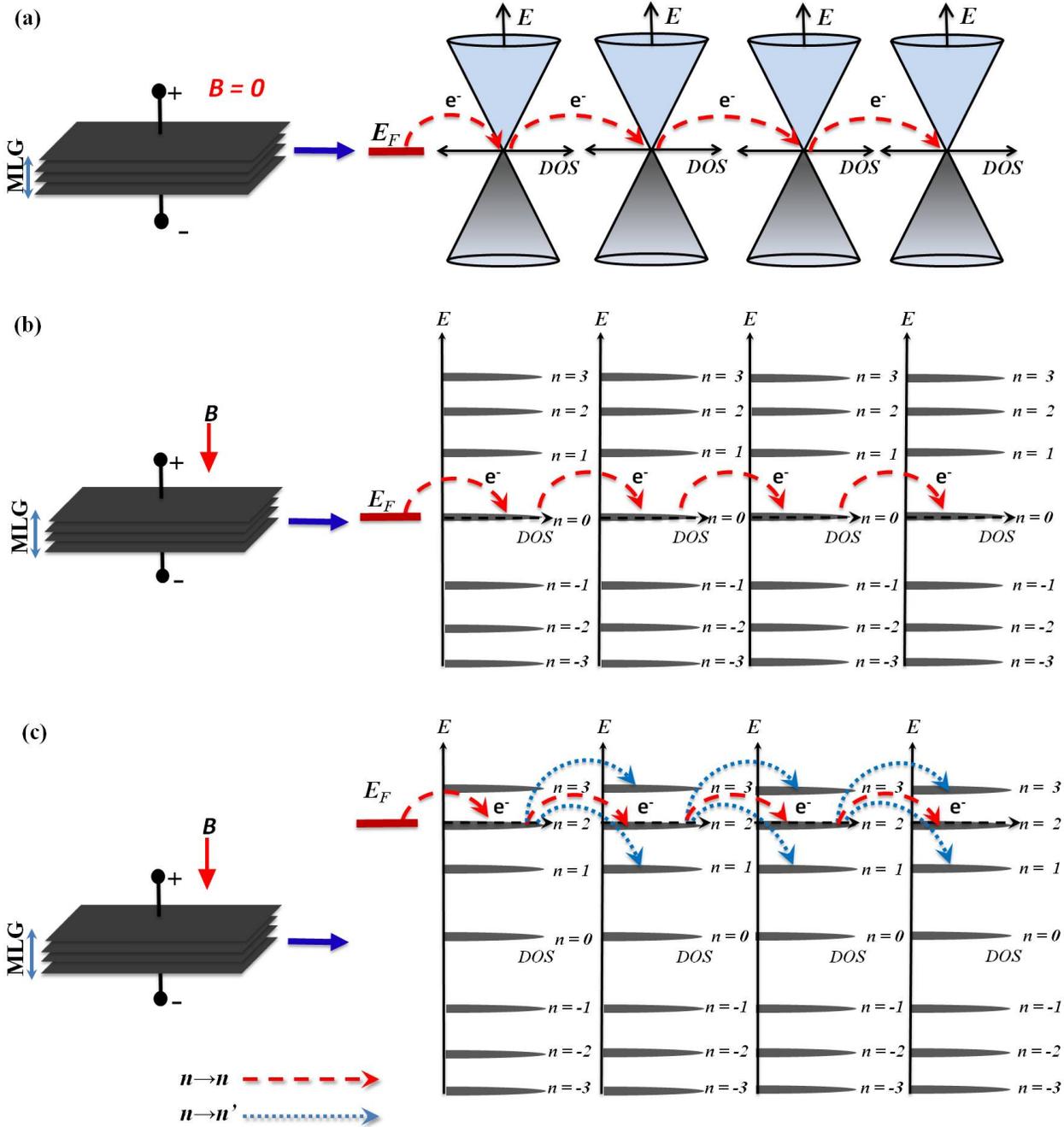

**Figure 1. Schematic description of the ILMR effect.** (a) Applied magnetic field ($B$) = 0 and an out of plane electrical bias drives the interlayer current. Weak interlayer coupling ensures that Dirac cone dispersion is preserved for individual layers. Carrier transport occurs via tunneling between the states in the vicinity of the (quasi) Fermi level $E_F$ (or Dirac point). Due to lack of available states (and hence available carriers) and weak interlayer coupling, interlayer current is weak. (b) Out of plane magnetic field is applied ($B \neq 0$) and a zero mode Landau level ($n = 0$) forms at the Dirac point, which coincides with the quasi Fermi level $E_F$. Each Landau level has a finite broadening due to disorder and thermal effects. Density of states (DOS) and carrier concentration of zero mode is proportional to $B$. Since interlayer transport occurs via zero mode, large interlayer current is observed due to large number of carriers participating from each layer. (c) Inter Landau Level mixing effect (dotted lines, blue), which is dominant when inter Landau level separation is small (such as small $B$, large broadening etc.). Dashed lines (red) show typical ILMR mechanism, without any mixing. In this example, $E_F$ is located away from the Dirac point.

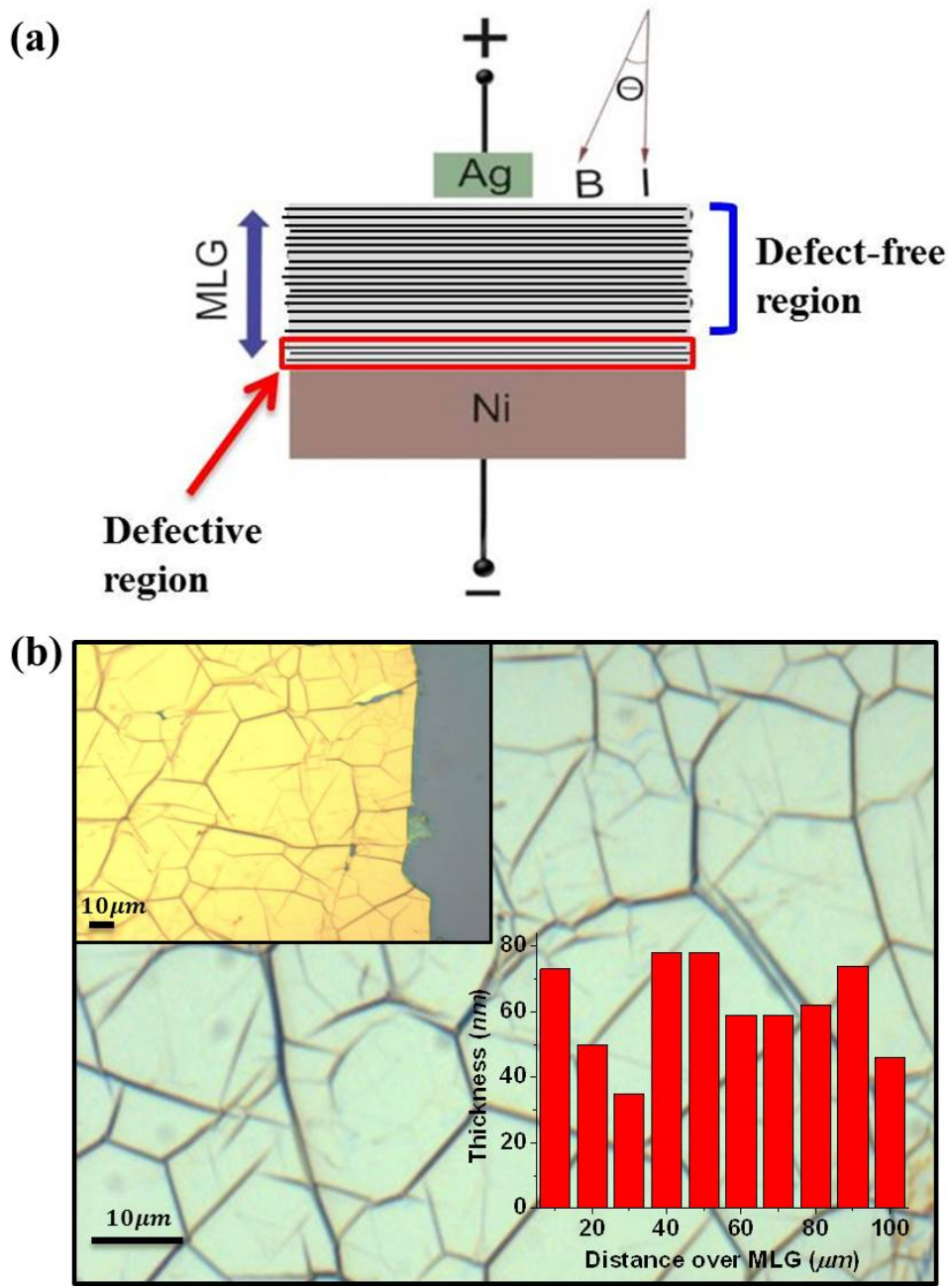

**Figure 2. Device schematic and optical images of as-grown and transferred MLG.** (a) Device structure and measurement geometry. The "tilt angle" $\theta$ is measured with respect to the out of plane direction. MLG grown on bottom Ni substrate consists of a "defective" region at the interface and a "defect-free" region at the top. The defect free region consists of weakly coupled graphene layers. Silver paste contact is placed at the centre of the top MLG surface to achieve uniform current distribution. (b) (*main image*) Optical micrograph of as-grown MLG on Ni. Top left *inset* shows the transferred MLG on SiO$_2$/Si. The histogram in the main image shows typical thickness distribution in the wrinkle free areas of transferred MLG. Average MLG thickness in the wrinkle free area is ~ 60 nm. In the optical images, the dark lines represent wrinkles (or regions of larger thickness) in the MLG layer.

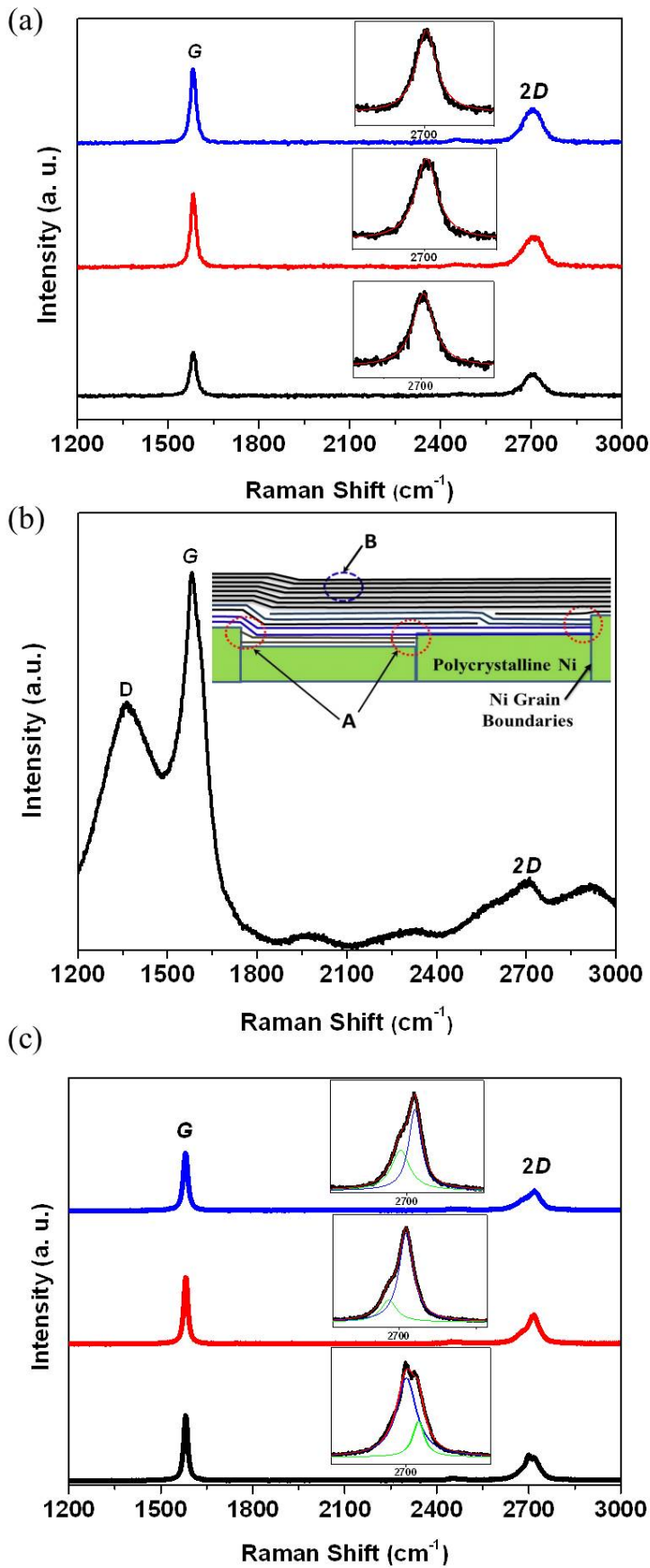

**Figure 3. Raman studies on as-grown and transferred MLG.** (a) Raman spectra from three representative areas of as-grown MLG on Ni (i.e. *before* transfer). The top plot (blue) is most commonly observed (~ 80% area). The 2*D* band is symmetric in all cases and can be fitted with single Lorentzian (insets). (b) Raman spectrum taken from the Ni/MLG interface after removing the Ni. Clear *D* peak is present, which confirms defective nature of this region. The inset shows schematic of graphene growth on Ni. Regions marked "A" (near Ni/MLG interface) are truncated by Ni grain boundaries and significant interfacial hybridization occurs in this region, which are the origins of defects in this region. But regions marked "B" (away from Ni/MLG interface) have continuous graphene layers covering the underlying layers. These layers are relatively defect free. (c) Typical Raman spectra of MLG (top layers) transferred on SiO$_2$/Si, from three representative regions. In all cases either splitting or shoulder in 2*D* band has been observed. No defect (*D*) peak has been observed in both as-grown and transferred MLG. Since the penetration depth of the Raman laser (2.33 eV) is ~ 50 nm and average sample thickness is ~ 60 nm, the Raman signal originates from the "defect free" region as described in Figure 2(a).

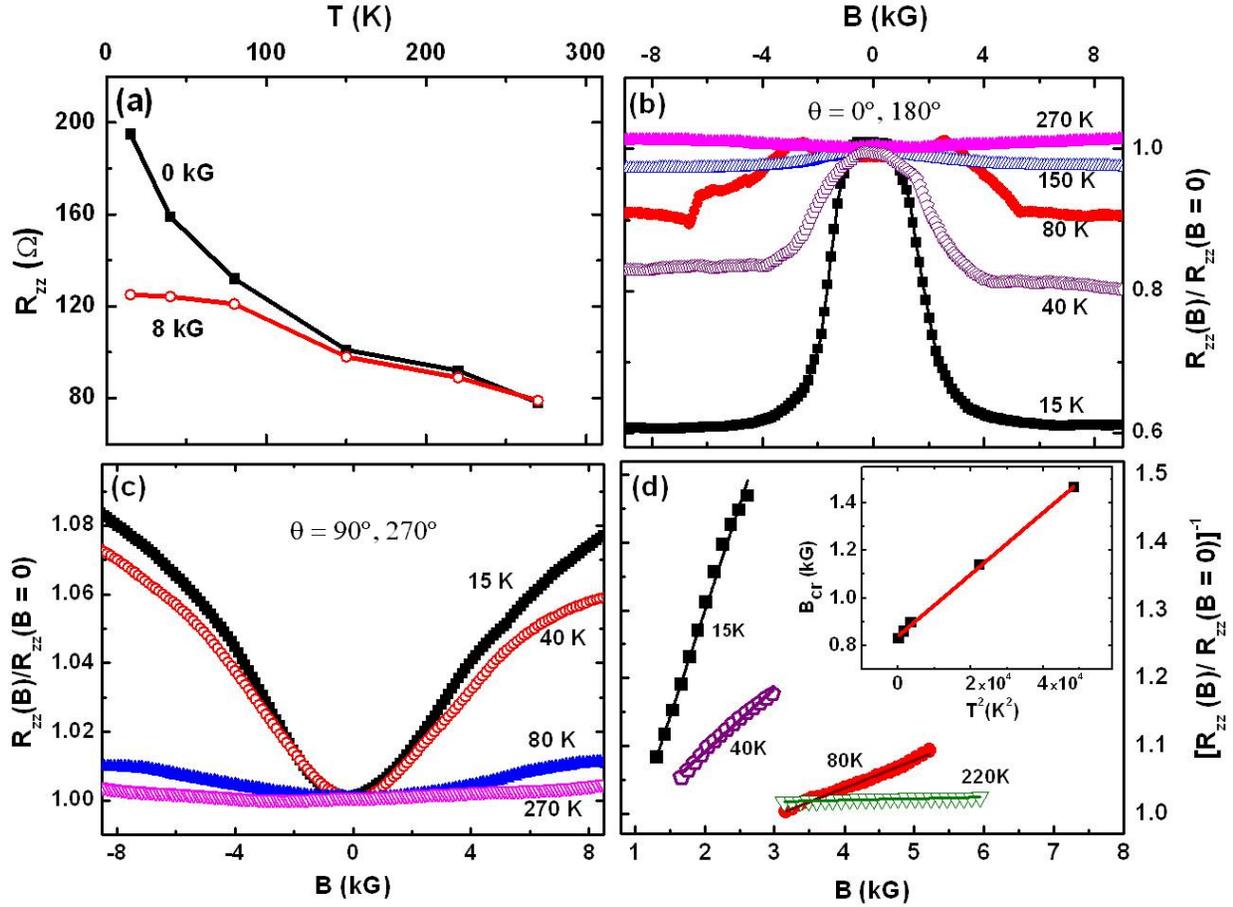

**Figure 4. CPP MR characterization for out of plane ($\theta$ = 0°, 180°) and in-plane ($\theta$ = 90°, 270°) magnetic fields.** (a) Temperature ($T$) dependence of CPP resistance $R_{zz}(T)$ in MLG/Ni samples at zero magnetic field and at 8kG (out of plane) over the temperature range 15 – 270 K. Insulating behaviour is observed at both temperatures, along with a magnetoresistance effect. (b) Normalized CPP resistance ($R_{zz}(B)/R_{zz}(B = 0)$) at various measurement temperatures for $\theta$ = 0°, 180°. A negative MR effect is observed. This effect weakens and MR curves broaden as temperature is increased. Critical field $B_{cr}$ is the field value at which device resistance starts to drop significantly. (c) Normalized MR ($R_{zz}(B)/R_{zz}(B = 0)$) at various measurement temperatures for $\theta$ = 90°, 270°. A positive MR is observed in this case. As before, MR effect weakens and MR characteristics broaden as temperature is increased. (d) Inverse of normalized CPP resistance ($R_{zz}(B)/R_{zz}(B = 0)$)$^{-1}$ as a function of out-of-plane magnetic field ($B$) in the range where negative MR is most prominent. Clear linear fit is observed in all cases. The inset shows variation of $B_{cr}$ as a function of $T^2$. A clear linear fit is observed.

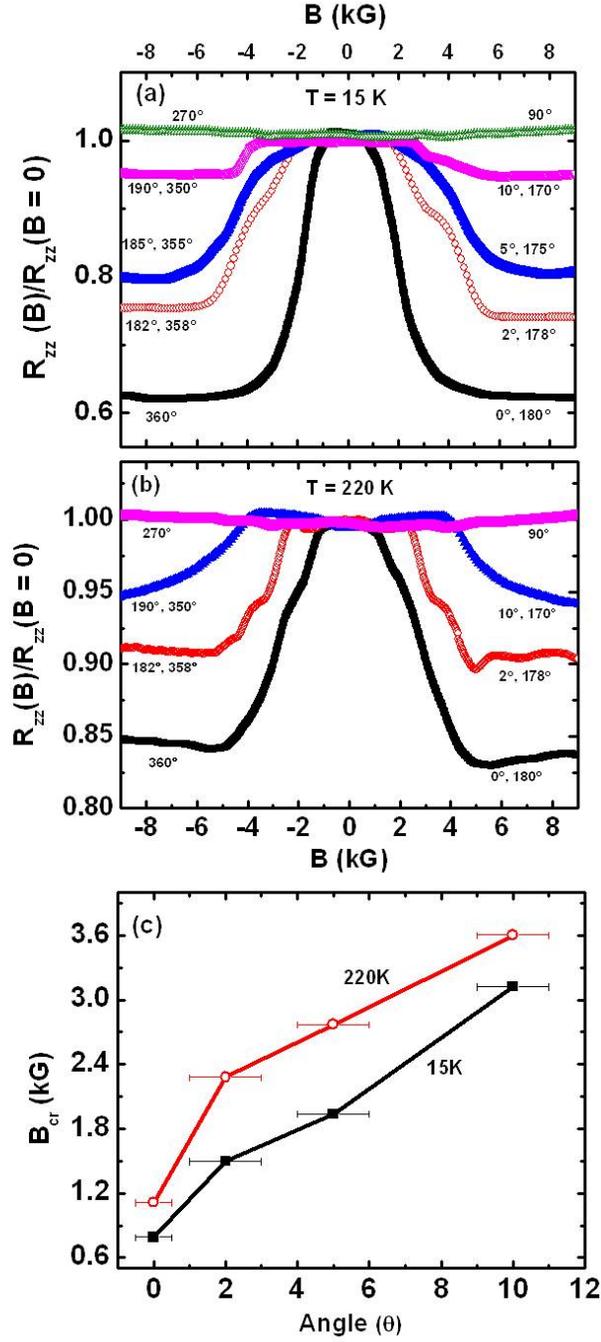

**Figure 5.** Angle dependence of CPP MR. (a), (b) Normalized CPP resistance ($R_{zz}(B)/R_{zz}(B = 0)$) of as-grown MLG on Ni at various orientations of the magnetic field ($\theta$) at two different temperatures (15 K and 220 K). The negative MR gradually decreases as tilt angle $\theta$ is increased. (c) Critical field ($B_{cr}$) as a function of $\theta$ at 15K and 220K. Critical field is higher at higher temperature and increases with the tilt angle.

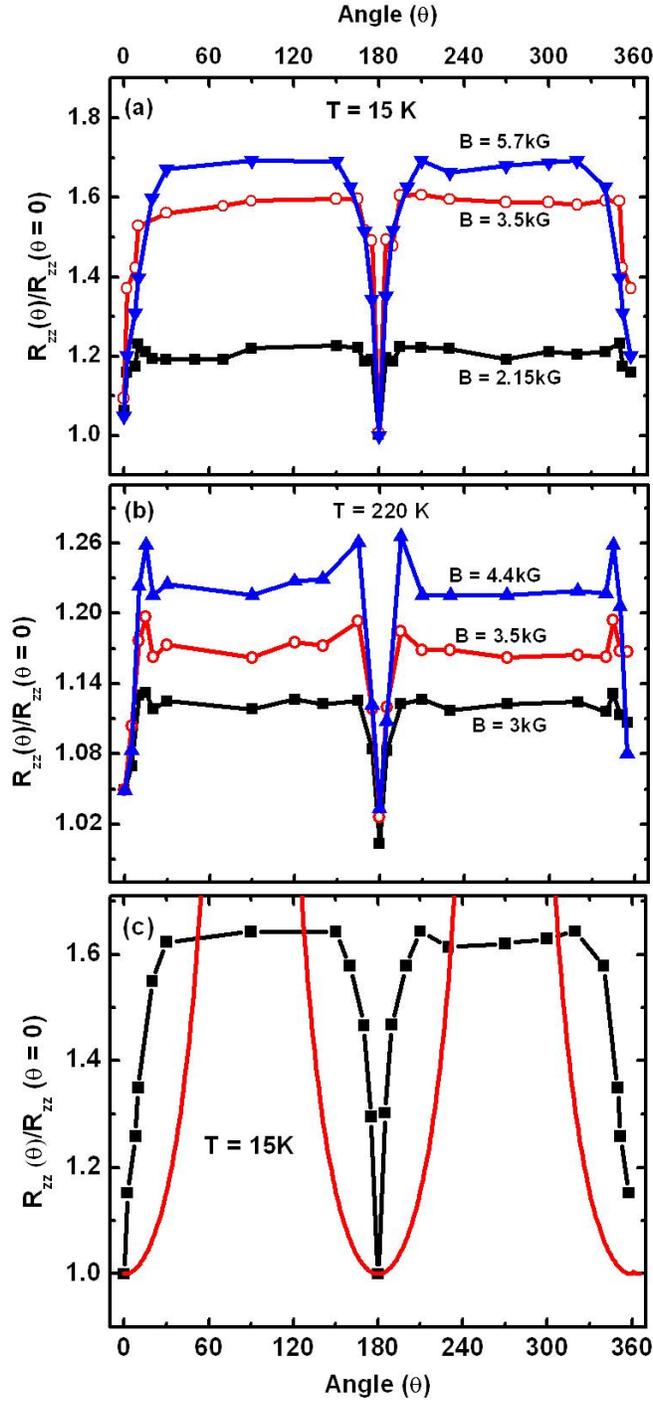

**Figure 6. Angular dependence of CPP resistance $R_{zz}(\theta)$ at various field strengths.** (a), (b) Data at 15K and 220K respectively. Three different field values are chosen, which are higher than $B_{cr}$. For a given field strength ($B$), device resistance at angle $\theta$ is normalized by the resistance value at $\theta = 0^o$, which also coincides with the resistance value at $\theta = 180^o$. As the tilt angle is increased with respect to $0^o$ (or $180^o$), device resistance increases and ultimately saturates for $\theta > 10^o$. (c) Experimental data (black line with data points) shows sharper angular dependence compared to theory (red, smooth curve). The data and the fit correspond to $B = 5.7$ kG. The theoretical curve diverges at $\theta = 90^o$, $270^o$ as discussed in the text.

# Supplementary Information.

## 1. In-plane Electrical Characterization of Transferred MLG – Sheet Resistance and Contact Resistance Measurements.

In the main paper, we reported characterization of as-grown and transferred MLG (both surfaces) using Raman spectroscopy (Figure 3). This method is widely used for characterization of graphitic nanostructures (graphene, graphite, carbon nanotubes etc.) due to its non-destructive nature and the wealth of information that can be obtained from such measurements[1,2]. In the main paper, we have shown that our as-grown samples (on Ni, not transferred) typically show formation of prominent hexagonal lattice of carbon atoms (strong *G* peak), weak interlayer coupling (symmetric 2*D* band), and absence of any defect (absence of *D* band). After transferring on $SiO_2$ substrate, 2*D* Raman band becomes distorted, indicating loss of weak interlayer coupling as a result of the transfer process. Such change is not surprising since there are several recent reports that have unearthed various non-idealities of the transfer process[3,4]. Nevertheless, to further characterize our samples, we have performed in-plane electrical measurements on the transferred specimens. Details of these measurements are described below.

Sheet resistance ($R_s$) of the transferred MLG has been measured using two methods. First, a "Transfer length (TLM) method" has been employed, from which sheet resistance ($R_s$) and contact resistance ($R_c$) have been evaluated. The typical device geometry is shown in **Figure S1(a)**. The electrical contacts used in TLM measurements are labeled as A, B, C, D in **Figure S1(a)**. From the TLM measurement (**Figure S1(b)**), $R_s$ is < 100 Ω/□ and $R_c$ is < 10 Ω within temperature range of 10K – 290K. **Table S1** lists the typical $R_s$ and $R_c$ values at two different sample temperatures.

Next, a van der Pauw geometry has been used to extract $R_s$ following a standard procedure[5]. Typical measurement configuration is shown in **Figure S1(a)** and the electrical contacts for this measurement are labeled as 1, 2, 3 and 4. At $T = 30K$, $R_s$ = 87.5Ω/□, whereas at 80K and 200K $R_s$ takes values 78 Ω/□ and 50.7 Ω/□ respectively. These values are consistent with those extracted by the TLM method (see **Table S1**) and give us confidence about the reliability of these numbers.

We note that such values of $R_s$ are typical for CVD grown MLG of similar thickness (on Ni), and similar values have been reported by several groups in the past[6,7]. Thus, electrical quality of our Ni-grown MLG samples is on a par with those reported in literature. Note that such low $R_s$ films are often used as flexible, conductive and (semi-)transparent electrodes in flexible optoelectronic applications as ITO replacement[6,7].

**Table S1.**

| Temperature (T) | $R_c$ (TLM method) | $R_s$ (TLM method) | $R_s$ (van der Pauw method) |
|---|---|---|---|
| 30K | 7.9 Ω | 98.7 Ω/□ | 87.5 Ω/□ |
| 200K | 2.9 Ω | 36.7 Ω/□ | 50.7 Ω/□ |

The contact resistance $R_c$ is also an order of magnitude lower than the CPP resistance values reported in the main paper. Thus Ag/MLG contact resistance does not play an important role in the observed MR.

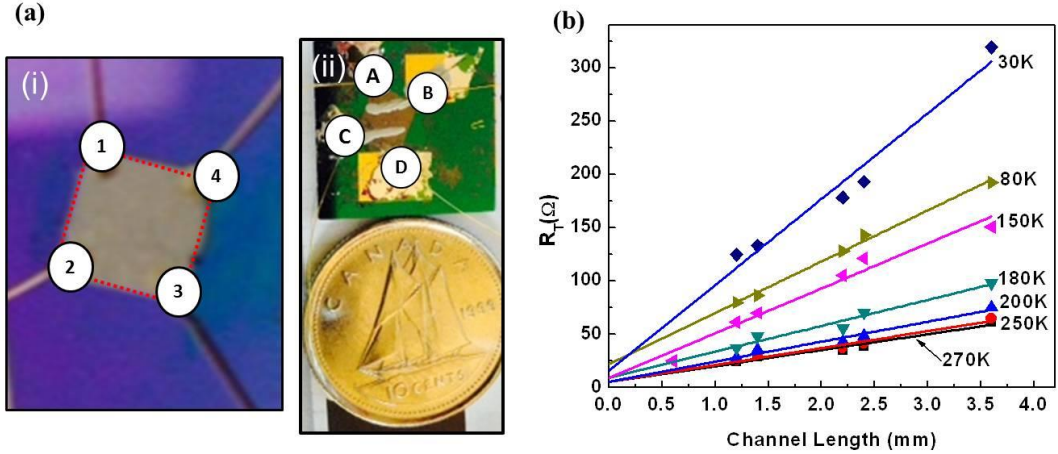

**Figure S1.** Electrical characterization of MLG transferred on SiO$_2$/Si. (a), (*i*) Typical van der Pauw geometry and (*ii*) TLM geometry. (b) TLM data, from which Ag/MLG contact resistance ($R_c$) and MLG sheet resistance ($R_s$) have been extracted.

**2. In-plane MR Characterization of Transferred MLG – Weak Localization Effect.**

As discussed in the main paper, graphene layers near the top surface are almost defect-free, since no defect-induced peak has been observed in the Raman spectrum. This is consistent with prior work on CVD-grown MLG on Ni, which noticed low defect content in such systems[7]. To further test this point, we performed temperature-dependent in-plane magnetoresistance measurements on these samples (**Figure S2(a)**). It is well known that the presence of grain boundaries and defects leads to weak-localization effect due to scattering of carrier wave functions[8]. However, no such effect has been observed in our Ni-grown MLG samples (after transferring on SiO$_2$, **Figure S2(a)**), which is consistent with the Raman data (Figure 3(a)) that does not indicate any presence of defects or scattering centers.

On the other hand, we have measured in-plane magnetoresistance of Cu-grown samples (**Figure S2(b)**), which shows pronounced weak localization effect. This is consistent with observation of strong defect peaks in the Raman spectra of Cu-grown samples (**Figure S2(c)**), which originate from the grain boundaries. Thus the top layers of our Ni-grown samples indeed have very low defect content.

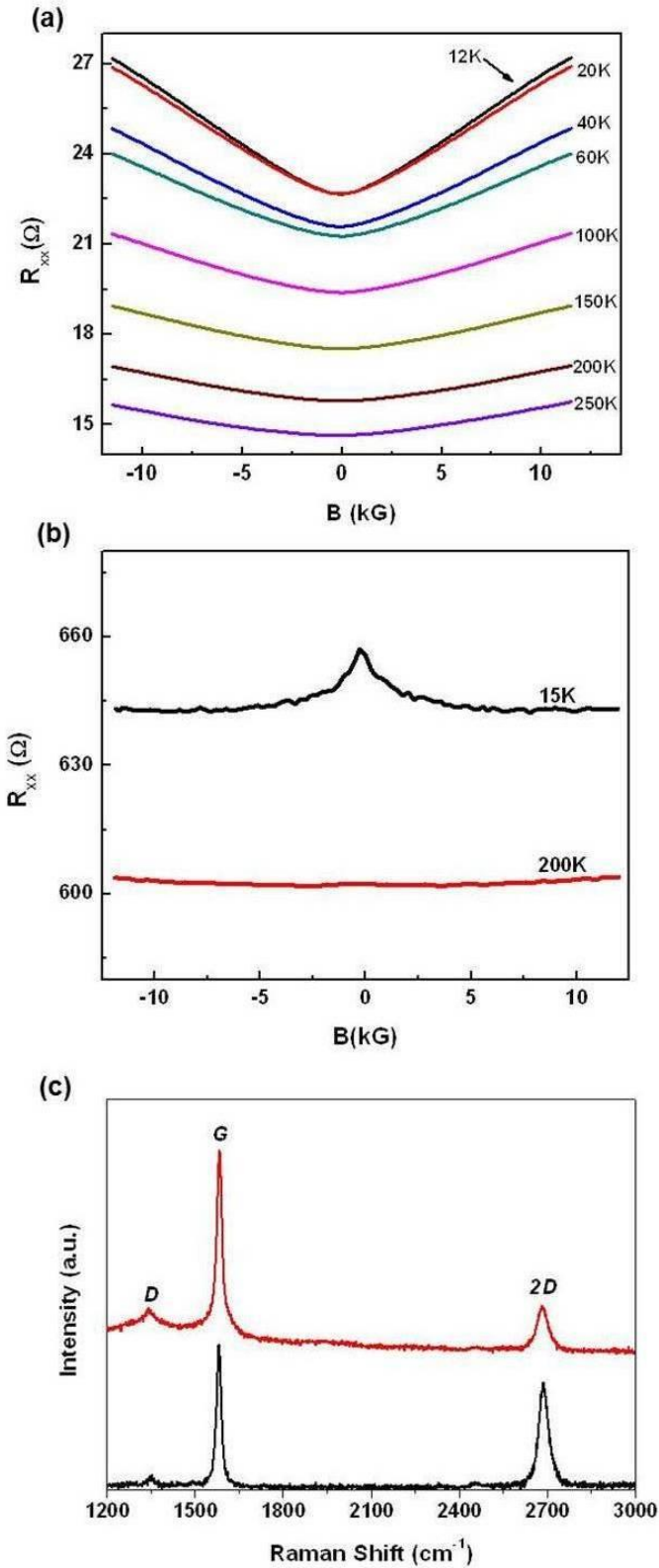

**Figure S2.** In-plane MR characterizations. (a) In-plane MR from MLG (top surface), transferred on SiO$_2$/Si, at various temperatures. Positive MR has been observed at all temperatures. No negative MR due to weak localization effect has been observed near $B = 0$. (b) In-plane MR from 8-layer transferred graphene (Cu-grown). Clear negative MR due to weak localization effect has been observed near $B = 0$. As expected, this effect disappears as temperature is increased. (c) Typical Raman data from 8-layer, Cu-grown graphene. Defect induced Raman peak ($D$-peak) is present, which originates from the grain boundaries. Weak localization effect observed in (b) is due to these grain boundaries.

## 3. Shubnikov-deHaas Oscillations in the in-plane MR Measurements.

The occurrence of ILMR as reported in the main paper relies on formation of distinct Landau levels in presence of an out-of-plane (i.e. parallel to the *c* axis) magnetic field ($B_z$). Due to in-plane scattering, Landau levels are broadened ($E_{LL} \pm \Gamma/2$) and for small values of magnetic field, inter-Landau level separation ($\Delta_{LL}(B_z)$) is small. As a result, for small magnetic field values Landau levels are not distinct ($\Delta_{LL}(B_z) < \Gamma$). However, as magnetic field strength is increased, inter-Landau level separation increases, eventually resulting in distinct Landau levels (($\Delta_{LL}(B_z)) > \Gamma$). Formation of distinct Landau levels leads to oscillations in in-plane resistance at higher $B_z$ values, which is commonly known as Shubhnikov-deHaas oscillations[9]. Prior works on various graphitic systems have reported observation of such oscillations for $B_z \geq 0.6T$[10–12], and the oscillations increase in amplitude as magnetic field is increased. This occurs because higher magnetic field leads to larger separation between the neighboring Landau levels. At the same time, oscillation amplitude decreases with increasing temperature because at higher temperature scattering induced broadening is larger due to thermal excitations.

To explore if Landau levels are formed within our measurement range of ±1 T, we have performed in plane magnetoresistance measurements on our transferred samples (CVD grown on Ni and subsequently transferred on $SiO_2$). **Figure S2(a)** shows the raw data (solid curves) at various measurement temperatures. To explore the presence of any underlying oscillation in the measured field range, we have fitted the experimental curves by monotonic backgrounds (**Figure S3(a)**). **Figure S3(b)** shows the residues after subtracting the background from the experimental data. A clear oscillatory behavior has been observed within our measurement range of ±1 T. As expected, the oscillation amplitude increases as field is increased and the oscillation amplitude is weakened as temperature is increased. However, phase and periodicity of oscillation remain almost unchanged. Oscillations have been detected up to $T = 250K$. Thus, distinct Landau level formation takes place within our measurement range of ±1 T, which is consistent with prior experiments on graphitic specimens[10–12].

Using the SdH data, carrier concentration per layer is estimated[9] to be ~ $10^{10}/cm^2$. Since the observed ILMR effect (main paper) manifests at ~ 2 kG, we can estimate the number of occupied Landau levels at this field value. Using the standard formula[9] (number of Landau levels = $(n_S)_{1L}/[2eB/h]$), we find that ~ 1-2 Landau levels are occupied. Thus, our devices operate very close to the so-called "quantum limit".

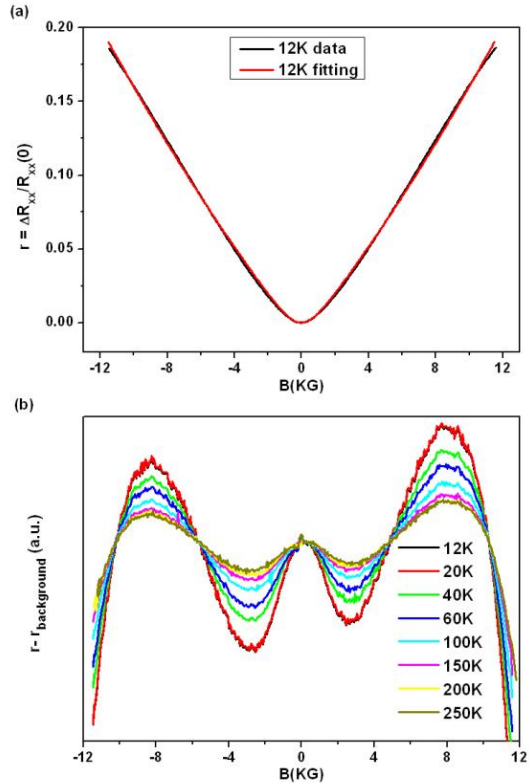

**Figure S3.** Shubnikov-deHaas oscillations in in-plane MR of Ni-grown MLG transferred on $SiO_2$/Si. (a) Typical MR plot at 12K (black line) and smooth monotonic background (red line). (b) MR at various temperatures after removing the smooth background. MR oscillations have been observed, which increase in amplitude as *B* is increased. Also, oscillation amplitude decreases as *T* is increased.